\documentclass[12pt]{article}
\usepackage[hyperfootnotes=false]{hyperref}
  \usepackage{graphicx}
  \newlength{\abstractwidth}
  \renewcommand{\thefootnote}{\fnsymbol{footnote}}
  \renewcommand{\thanks}[1]{\footnote{#1}} 
  \newcommand{\starttext}{
  \setcounter{footnote}{0}
  \renewcommand{\thefootnote}{\arabic{footnote}}}
  \renewcommand{\theequation}{\thesection.\arabic{equation}}
  \newcommand{\be}{\begin{equation}}
  \newcommand{\bea}{\begin{eqnarray}}
  \newcommand{\eea}{\end{eqnarray}}
  \newcommand{\beq}{\begin{equation}}
  \newcommand{\ee}{\end{equation}}
  \newcommand{\eeq}{\end{equation}}

  \def\ba{\begin{eqnarray}}
  \def\ea{\end{eqnarray}}

  \def\12{{1 \over 2}}
  \def\eq{&=&}

\def\R{\bar{r}}

  \def\simleq{\; \raise0.3ex\hbox{$<$\kern-0.75em
      \raise-1.1ex\hbox{$\sim$}}\; }
   \def\simgeq{\; \raise0.3ex\hbox{$>$\kern-0.75em
      \raise-1.1ex\hbox{$\sim$}}\; }

\def\ba{\bf{a}}

  \def\h3{{\cal{H}}_3}

  \def\A{{\cal{A}}}
    \def\R{{\cal{R}}}

\def\O3{\Omega_3}
\def\o2{\Omega_2}
\def\o{\omega}
\def\bi{\begin{itemize} }

\def\ei{\end{itemize} }

\begin{document}
  \renewcommand{\theequation}{\thesection.\arabic{equation}}
  \begin{titlepage}
  \rightline{SU-ITP-10/34}
  \bigskip

  \bigskip\bigskip\bigskip\bigskip

    \centerline{\Large \bf {Addendum to Fast Scramblers}}

  \bigskip \bigskip

  \bigskip\bigskip
  \bigskip\bigskip

\begin{center}
  {{Leonard Susskind  }}
  \bigskip

\bigskip
Stanford Institute for Theoretical Physics and  Department of Physics, Stanford University\\
Stanford, CA 94305-4060, USA \\

\vspace{2cm}
  \end{center}

  \bigskip\bigskip
  \begin{abstract}

This paper is an addendum to \cite{Sekino:2008he} in which I point out that both de Sitter space and Rindler space are fast scramblers. This fact supports the idea that the holographic description of a causal patch of de Sitter space may be a matrix quantum mechanics at finite temperature \cite{Banks:2006rx}. The same can be said of Rindler space. Some qualitative features of these spaces can be understood from the matrix description.

  \medskip
  \noindent
  \end{abstract}

  \end{titlepage}
  \starttext \baselineskip=17.63pt \setcounter{footnote}{0}

\bigskip
\setcounter{equation}{0}
\section{Fast Scramblers}

The scrambling time $t^\ast$ (as defined in \cite{Sekino:2008he}) for a system at temperature $T$ is the time that it takes for a perturbation, such as the addition of a particle to the system, to become so thoroughly mixed among the degrees of freedom, that the information it contains can only be retrieved by examining a fraction $\sim 1$ of all the degrees of freedom \cite{Page:1993df}. It is obvious that for ordinary local systems, such as quantum field theories or fluids, that the scrambling time must be at least as long as  the ordinary diffusion time, $t_D$, i.e., the time for a particle to diffuse over the entire size of the system.

Consider the diffusion of a particle in a fairly dense gas or liquid of molecules. Assume that the separation of the molecules $l_0$ is of the same order as their size so that there is only one length scale in the problem. Let the dimension of space be $d$. It is a simple exercise in kinetic theory to show that the time scale $t_D$ for a particle to diffuse over the entire system satisfies
\be
t_D T \approx N^{2\over d} l_0 \bar{p}
\label{1.1}
\ee
where $N$ is the number of molecules, $T$ is the temperature, and $\bar{p}$ is the average momentum of a molecule in thermal equilibrium.

Ordinary classical kinetic theory applies when the de Broglie wavelength, $\lambda =2\pi \hbar/p,$ of the molecules is much less than $l_0.$ When the wavelength  becomes of order the inter-particle spacing, $\lambda \sim l_0,$ we enter into the regime of strongly correlated quantum fluids and in this case \ref{1.1} becomes
\be
t_D T\approx N^{2\over d} \hbar
\label{1.2}
\ee

For the moment I will assume that for the case of strongly correlated quantum systems  the diffusion and scrambling times are of the same order, $t^\ast \sim t_D.$

If  the total entropy of the gas is of order the number of molecules  we many also write
\be
t^\ast T \approx S^{2\over d} \hbar
\label{1.3}
\ee

The range of validity of this formula is fairly broad: it requires the temperature to be high enough that the entropy is proportional to the number of particles, and low enough so that the de Broglie wavelength is comparable to the inter-particle spacing.

 Note that for a given entropy, the time-scale $t^\ast$ decreases with increasing dimension. One might think that in the real world $t^\ast T$ never grows more slowly than $S^{2\over 3}$. Thus it is remarkable that for a black hole, the scrambling time satisfies \cite{Hayden:2007cs} \cite{Sekino:2008he}
\be
t^\ast T \approx \hbar \   \log{S}
\label{1.4}
\ee
In effect the degrees of freedom describing a black hole behave as an infinite dimensional system in which every degree of freedom is directly coupled to every other, so that information can diffuse maximally rapidly. In \cite{Sekino:2008he} Sekino and I referred to such unusual systems as ``fast scramblers."

It was also noted in \cite{Sekino:2008he} that the origin of such behavior in cases where the microscopic description of black holes is known, is the fact that the degrees of freedom are very delocalized matrix degrees of freedom as in Matrix theory\footnote{It is likely that matrix systems are not the unique fast scramblers. There are black hole constructions in string theory which seem to require more exotic elements. I thank Tom Banks and Eva Silverstein for discussions on this matter. }. The reason is that in matrix quantum mechanics every matrix element is directly coupled to every other matrix element by the quartic couplings, typically of the form
\be
-Tr [X^i , X^j][X^i , X^j]
\label{1.5}
\ee
In this formula the $X^i$ represent a collection of $N\times N$ matrices. It is a simple matter to check that every matrix element $X^i_{mn}$ is coupled to every other matrix element $X^j_{rs}$ by some term in the trace. This is in contrast to a lattice Hamiltonian in which every site is coupled to $\sim d$ nearest neighbors.

On the black hole side, the argument for why the diffusion time is so fast  was essentially classical, except for the introduction of a stretched horizon about one Planck (or string) length thick. Call the stretched horizon thickness $l_s$. One can simply drop a charged particle into the black hole and calculate how the charge density diffuses over the stretched horizon. The time that it takes to spread a distance $L$ is of order  $MG \log {L/l_s} $ \cite{Susskind:2005js} \cite{Hayden:2007cs}. This type of formula is very general for any horizon which is locally Rindler-like. Setting $L=2MG$, $S=4\pi M^2 G,$ and $1/T= 8\pi MG,$ for a Schwarzschild black hole shows that the diffusion time satisfies \ref{1.3}. The fact that the charge diffuses so rapidly does not depend on the details of the Schwarzschild metric but only on the Rindler-like character of the horizon.

\subsection{A Conjectured Bound}
Equation \ref{1.3} \ should really  be interpreted as a inequality rather than an equality since all we really know is that the scrambling time cannot be smaller than the diffusion time. Thus for $\lambda \sim l_0,$
\be
{t^\ast T  \over  S^{2\over d}}   \geq c \hbar
\label{1.6}
\ee
where $c$ is some constant.

In fact we would like to conjecture that \ref{1.6} is a universal bound on the behavior of $t^\ast.$ It is clear from the derivation that for high temperatures the diffusion time is longer by the factor $ l_0  /  \lambda$ than \ref{1.3}.  So one cannot beat the bound by raising the temperature. On the other hand at very low temperatures the properties of fluids are described by weakly coupled quasi-particles which are not efficient at scrambling information. There are of course many systems which are very far from saturating the bound.

If the bound \ref{1.6} is correct then one may say that systems that saturate it are the fastest scramblers in $d$-dimensions. But we will reserve the term fast scramblers for those systems that scramble in logarithmic time.

\setcounter{equation}{0}
\section{de Sitter Space is a Fast Scrambler}

The purpose of this paper is very modest. It is mainly to point out that de Sitter space and Rindler space are fast scramblers and to consider the possibility that the origin of the  fast-scrambling property is matrix quantum mechanics or something quite like it.
  I will not be able to answer any of the deep questions concerning observer complementarity, or the full symmetry group of de Sitter space that takes one causal patch to another. Nor will I consider detailed string constructions.  The concern will be with  the number of  degrees of freedom that it takes to describe these spaces, and especially, how must they be connected by direct interactions.

The radius of de Sitter space is inverse to the Hubble constant $H,$
$$
\R=1/H
$$

Consider a static patch of de Sitter space with metric,
\be
ds^2 = -(1-H^2r^2)dt^2 +  (1-H^2r^2)^{-1} dr^2 +r^2 d\Omega_2^2.
\label{2.1}
\ee

The shell at $r=\R$ is a Rindler-like horizon. As in the black hole case the mathematical horizon can be stretched or thickened by amount $l_s.$
Suppose that a charged particle passes through the horizon. Once again the charge will diffuse over the stretched horizon. The result is the same as for a black hole: the charge diffuses to distance $L$ after time $\R \log {L/l_s}$. Letting
\bea
L \eq \R \cr
T  \eq  1/(2\pi \R) \cr
 S \eq \pi R^2 /G
 \label{2.2}
\eea
again gives the fast-scrambler formula
\ref{1.3}
which can also be written in the form
\be
t^{\ast}T = \R\log{(\R/ l_s)}
\label{2.3}
\ee

From the fact that the de Sitter horizon is a fast scrambler we can conclude that it cannot be described by any kind of two-dimensional field theory.

To see just how fast \ref{2.3} is, let us  consider how long it would take to scramble the information on the de Sitter horizon of our universe. First consider the incorrect case of a conventional local system---a 2-D lattice or a QFT.   In that case we would use the formula \ref{1.3} with  $d=2,$  $S \sim 10^{120},$ and $T=10^{-60}$ in Planck units. The result would be about $t^{\ast} \sim 10^{130}$ years. On the other hand, using \ref{2.3} gives the vastly smaller answer $t^{\ast} \sim 10^{12}$ years.

Given that de Sitter space is a fast scrambler, it seems likely that a dual description based on matrix quantum mechanics should exist, at least as an approximation (we will return to the question of precision in Section 6.).
The idea that de Sitter space can be described by in terms of matrix quantum mechanics  has been long advocated by Banks and others, \cite{Banks:2006rx}.

\setcounter{equation}{0}
\section{Rindler Space is a Fast Scrambler}

There are various ways to approach flat space from geometries that have known or suspected dual descriptions. The oldest is BFSS Matrix Theory \cite{Banks:1996vh}. ADS/CFT can also be studied in a limit in which the
curvature goes to zero. The static patch of de Sitter space can be used in a number of ways to approach flat space. The most obvious is to sit at the center of the static patch and let $\R$ go to infinity. But there is also a second way that we will consider here. One can place an observer at
a fixed proper distance from the horizon and then let $\R$ go to infinity. In that case the observer
is uniformly accelerated and in the limit the spacetime tends to Rindler space.

Rindler space is just flat spacetime as seen by a uniformly accelerated observer. In terms of the Cartesian coordinates $(x^0, \ x^1, \ x^2,  \ x^3)$ the Rindler coordinates are $(\omega, \ \rho, \ x^1, \ x^2)$
where $\rho, \ \omega $ are defined by,
\bea
x^0 \eq \rho \ \sinh{\omega} \cr
x^3 \eq \rho \ \cosh{\omega}
\label{3.1}
\eea

The metric of Rindler space is given by
\be
ds^2 = dx^idx^i + d\rho^2 - \rho^2 d\omega^2
\label{3.2}
\ee
where $x^i$ are the coordinates of the 2-dimensional plane, $\rho$ is the proper distance from the horizon, and $\omega$ is the dimensionless Rindler time.

The Rindler Hamiltonian is dimensionless, being the generator of $\omega$-translations.
Similarly the Rindler temperature is dimensionless and has the universal value
\be
T_r = {1 \over 2 \pi }.
\label{3.3}
\ee

Rindler space is the limit of a fast scrambler. A charge dropped onto the horizon spreads
with a profile that grows  exponentially with $\omega$ (see \cite{Sekino:2008he} ) so that it covers a region of size
\be
\Delta x = l_s e^{\omega}
\label{3.4}
\ee
after time $\omega.$

The entropy of Rindler space is infinite but we can regulate it by bounding the $x$-plane to have area $\A$ in Planck units. In that case the charge distribution covers the entire regulated plane in Rindler time
\be
\omega^{\ast} = \log{\A/l_s^2}
\label{3.5}
\ee

This, together with the fact that the temperature is order unity, shows that Rindler space is a fast scrambler, or more exactly the limit of a fast scrambler. Obviously, the matrix theory describing the regulated region should be in terms of $N\times N$ matrices of size
\be
N^2 \sim \A.
\label{3.6}
\ee
In the limit of infinite area Rindler space must be a theory of infinite matrices.

\setcounter{equation}{0}
\section{A Naive Matrix-Rindler Model}

The manifest symmetries of Rindler space are time ($\omega$) translation and $x$ translation. I will make a guess about the matrix degrees of freedom, namely that they  include matrices $X^i$ which are the matrix analogs of the coordinates $x^i.$
One can think of the $x's$ as the eigenvalues of the $X$-matrices as in Matrix Theory \cite{Banks:1996vh}. In addition fermionic degrees of freedom and  some form of supersymmetry, are needed to keep the theory stable \cite{VanRaamsdonk:2001jd}. Compactification of extra dimensions will also introduce additional structure.
In this paper these ``details" will be ignored although at times the fermionic degrees of freedom and supersymmetry will be invoked.

The action of the matrix theory should reflect the manifest symmetries. In particular it should be invariant under ordinary two dimensional rotations and translations, Rotations act in the obvious way on the two matrices $X^1, \ X^2.$ The translations are defined by,
\be
X^i \to X^i + c^i I
\label{4.1}
\ee
where $I$ is the identity matrix and $c^i$ is any numerical 2-vector. The simplest candidate terms are a kinetic term (dot means derivative with respect to $\omega$.)
\be
K = Tr \dot{X^i}\dot{X^i}
\label{4.2}
\ee
and a potential term
\be
U = - Tr [X^1, X^2]^2
\label{4.3}
\ee

Thus, ignoring numerical constants, the action has the form
\be
A=\int  Tr \left\{{1 \over 2} \dot{X}^i   \dot{X}^i  + [X^1 , X^2]^2
\right\} \ d\omega
\label{4.4f}
\ee

Finally, the temperature is $T_r = 1/(2\pi).$

To connect Rindler space and the static patch of de Sitter space one can use the following correspondence principle as a guide:
\begin{itemize}
\item Identify the area of the horizon of de Sitter space $\R^2$ with the regulated area $\A$ of the Rindler horizon.

\item Identify the center of the static patch of de Sitter space, with the point $x=0, \ \rho =\R = H^{-1}.$ The size of the de Sitter space has been chosen so that that the corresponding points ( one in de Sitter space and the other  in Rindler space) are at the same distance from the respective horizons.

    \item  Consider some phenomenon taking place at a distance $\rho$ from the horizon of Rindler space. Assume the phenomenon occupies a region of the $x$ plane of size $\Delta x$. To describe it using the correspondence principle, we require the de Sitter space radius $\R$ to be bigger than both $\rho$ and $\Delta x.$

       In particular if $\Delta x < \rho$ then the most efficient description is to choose $\rho \sim \R.$

\end{itemize}

 \begin{figure}
\begin{center}
\includegraphics[width=18cm]{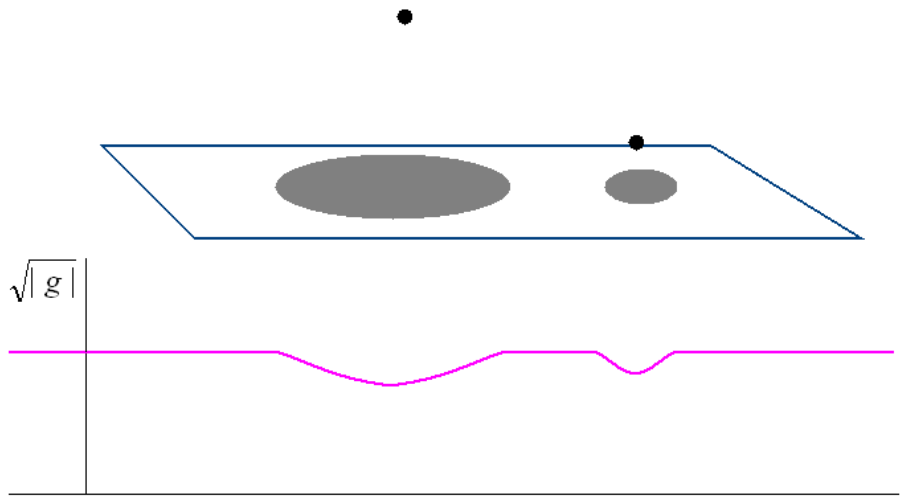}
\caption{Phenomena at distance $\R$ from the Rindler horizon correspond to the same phenomena at the center of a de Sitter static patch in which the de Sitter radius is $\R.$}
\label{1}
\end{center}
\end{figure}

Consider a massive object $M$ which is momentarily at rest at a distance $\R$ from the Rindler horizon localized near $x^i=0$. I will call it a black hole, but it could be any object with a long-range gravitational field given by the Schwarzschild metric.
If one follows it backward in time it will fall into the stretched horizon: if one follows it forward in time it will fall into the stretched horizon. (The same is true in de Sitter space unless the object is balanced at the center of the static patch.) The right way to think of the object is to imagine that it was shot out of the stretched horizon  by a thermal fluctuation. It  moved out to some maximum distance from the horizon,  then fell back in, and became scrambled among the horizon degrees of freedom.

Using the correspondence principle at the point where the object has reached its maximum distance---namely $\R$---from the horizon we can relate the configuration to a corresponding configuration of the same object, at the center of a de Sitter static patch of Hubble radius $\R.$

In de Sitter space the presence of the mass $M$ has an effect on the horizon: It shrinks the area of the horizon  by an amount,
\be
\Delta \A =  -GM\R.
\label{4.5}
\ee
Equivalently the entropy is diminished by amount
\be
\Delta S = - M\R
\label{4.6}
\ee

By the correspondence principle the same object, at distance $\R$ from the Rindler horizon, decreases the area of the Rindler horizon by \ref{4.6}.

I have used the correspondence principle mostly for brevity, but an actual calculation of a black hole in Rindler space is not difficult. Keep in mind that the spacetime is flat and that in cartesian coordinates the black hole is at rest at location $x^i = 0,  \ x^3 = \R$. To compute the effect of the black hole on the Rindler horizon is a matter of determining the horizon by following light-sheets from the remote future/past and large $x^3,$ back to the surface $x^0 = 0.$ Using the standard focusing equations one finds the same answer, i.e.,  \ref{4.5}, as by the correspondence argument. The argument is illustrated in Figure 2.
 \begin{figure}
\begin{center}
\includegraphics[width=12cm]{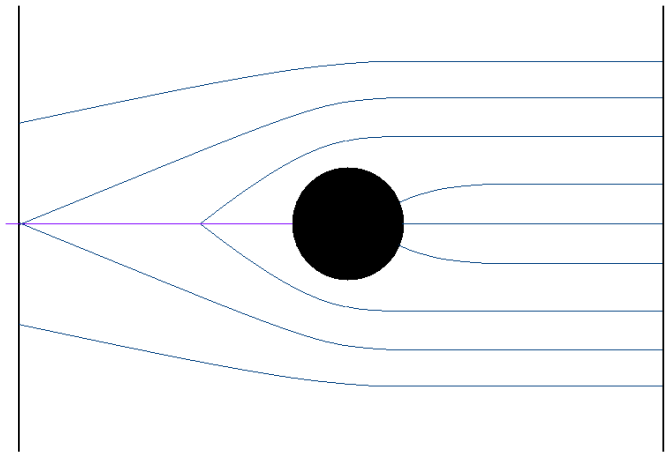}
\caption{The effect of a mass at $\R$ is to focus light rays from infinity and diminish the area of the horizon. }
\label{2}
\end{center}
\end{figure}

One can describe the perturbation on the  horizon in somewhat more detail. Consider the induced metric on the bifurcate horizon.
\be
ds^2 = g_{ij}dx^i dx^j
\label{4.7}
\ee
Far from $x=0$ the metric is flat: $g_{ij} \to \delta_{ij}$. The depleted area is given by
\be
\Delta \A = \int d^2 x \left\{\sqrt{|g|}  -1     \right\}
\label{4.8}
\ee
One finds not only that the depleted area is given by \ref{4.5} but that it is spread over a coordinate
area of order $\R^2.$ Thus the width of the region over which the entropy/area is depleted is a crude measure of the distance of the mass  from the horizon.

If the area of the horizon is regulated to be of order $\R^2$ in Planck units then the size of the matrices describing the regulated region is
\be
N\times N \sim \R \times \R.
\label{4.9}
\ee

The fact that an object  at $\rho$ affects the horizon out to distance $\Delta x = \rho$ can be expressed in another way: the object affects those degrees of freedom that account for an entropy of order $\rho^2.$ This allows us to add one more ingredient to
 the Rindler-de Sitter correspondence rule. It is the matrix analog of the above fact:
\begin{itemize}
\item The description of phenomena with $\rho < \rho_{max}$ and $\Delta x < \rho_{max}$ only involves a subgroup (of $SU(\infty)$) of size $N = \rho_{max}.$ In particular the excitation of mass at $\rho$ affects a $\rho \times \rho$ block with all other matrix elements unaffected.

\end{itemize}

Now let us consider how the configuration, including the mass $M,$ at distance $\rho$ is represented in the matrix theory. From the above rule we can regulate the size of the  matrices to be $N \times N$ with $N=\rho$.
The obvious idea familiar from Matrix Theory is to represent the configuration by block-diagonal $N\times N$ matrices,  containing a small block of size $m\times m,$ and a larger thermalized block of size $(N-m)\times (N-m)$. The degrees of freedom of the small block have been removed from the thermal ensemble and the off diagonal elements have been frozen into their ground state.

This corresponds with what we would do in de Sitter space. If the entropy of the de Sitter space is $\R^2,$ then the size of the matrices needed to describe all phenomena in the static patch is $N\times N,$ with $N = \R.$ To describe an object at the center of de Sitter space we would consider  similar block-diagonal matrices in which an $m\times m$ block is separated from the thermalized degrees of freedom. By separating the $m\times m$ block I mean that the off diagonal elements are put into their ground state.

After separating off the small block, the thermal degrees of freedom are  represented by $N-m$ dimensional matrices, and the thermal entropy has shrunk by an amount
\be
\Delta S = (N-m)^2 - N^2 = -2Nm.
\label{4.10}
\ee
Let us equate \ref{4.10} with \ref{4.6}. This gives
\be
M\R = Nm
\label{4.11}
\ee
Using \ref{4.9} and restoring the units allows us to compute the value of $m$ in terms of the mass $M$.
\be
Ml_p =m
\label{4.12}
\ee
In other words the size of the small block is determined by the mass of
the object in Planck units, and nothing else.
All of this is independent of the actual nature of the  object carrying mass $M$.
The details of the  object are encoded in the  state of the $m\times m$ block.

If  the massive object really is a black hole then we can use the above reasoning to give a derivation of its entropy. A black hole corresponds to a isolated thermal state of the $m\times m$ block, and therefore has entropy of order
$m^2.$ According to \ref{4.12} \  $m^2 = M^2 l_p^2,$ which is of exactly the right order to be the Bekenstein-Hawking entropy.

The location of the mass in the $x$ plane is straightforwardly represented by the $m\times m $ block. If the mass is centered at $x$
then the eigenvalues of the small block should also be centered at $x$. Let us henceforth place the object at $x=0.$

The location of the object in the radial direction is more subtle. If the mass is at distance $\rho$ from the horizon, then, as we have seen, the depletion within the horizon extends out to a distance $|x| \sim \R.$ An important question is how the distance $\rho$ is encoded in the state of the matrix theory.
A simple diagnostic is to consider the expression,
\be
F(r) = Tr \ e^{-X^iX^i \over r^2}.
\label{4.13}
\ee
This basically measures the number of eigenvalues in out to distance $r$. In the absence of the mass $M$
translation invariance would make $F(r)$  proportional to $r^2$. But the depletion of entropy from the region near $x=0$ will make $F(r)$ smaller for $r< \R$. Thus the function $F(r)$ contains the information about the distance of the mass from the horizon. This is shown in Figure 1.

The relation between the distance of the mass $M$ and the size of the affected region of the horizon is very reminiscent of anti de Sitter space. In that case the distance of an object to the UV boundary is reflected in the size of the affected region on the UV boundary. In the Rindler case the situation is both similar and different. The similarity is that the distance of the surface, in both cases determines range over which the surface is affected. But in the Rindler case the surface is not the UV boundary (which one might think of as $\rho = \infty,$ but instead it is the IR Rindler horizon at $\rho = 0.$

One point that may be of interest is that the depletion of the horizon entropy is related to the distance of the mass from the horizon in a manner that is identical to E. Verlinde's conjecture \cite{Verlinde:2010hp}, supporting his claim that at least in this context, gravity is an entropic force.

We can also consider forces between two objects at distance $\R$ from the horizon and separated by distance $\Delta x.$ In this case the $X$ matrices must contain three blocks, the two small blocks being of size $m\times m$ and $n \times n.$ $m$ and $n$ are related to the masses of the objects as in \ref{4.12}. The gravitational force between the masses comes from integrating out the $mn$ off diagonal elements which have very high energy if the separation $\Delta x$ is large.

With only the bosonic matrices $X$ the force between the objects in much too strong. The frequency of the off diagonal elements is of order $\Delta x,$ which would result in a zero-point energy of order $mn |\Delta x|.$ In other words the potential energy would grow linearly with distance \cite{VanRaamsdonk:2001jd}.

The same thing would happen in matrix theory where this undesirable effect is canceled by the supersymmetric fermionic degrees of freedom. Thus it would seem that the matrix theory of Rindler space must also contain fermions and probably supersymmetry. Banks and collaborators have emphasized the importance of supersymmetry in the holographic description of de Sitter space \cite{Banks:2006rx}.

\bigskip

The correspondence principle allows us to guess how a matrix theory of de Sitter space may look. The translation symmetry of Rindler space can be replaced by the $O(3)$ symmetry of a spherical horizon. The
two matrices $X^i, $ would be replaced by three, $N\times N$ matrices $(X^1, \ X^2, \ X^3)$ with a constraint,
\be
(X^1)^2 + (X^2)^2 + (X^3)^2 = \R^2
\label{4.14}
\ee
and the action by
\be
A=\int  Tr \left\{{1 \over 2} \dot{X}^i   \dot{X}^i  + [X^1 , X^2]^2 + [X^1 , X^3]^2
+ [X^3 , X^2]^2
\right\} \ d\omega
\label{4.15}
\ee

Assuming the entropy is proportional to $N^2$ we are led to set $N=\R.$ The dimensionless temperature, $T\R$ is $1/2\pi$ as in Rindler space. Banks, Fiol and Morrisey describe a more ambitious theory based on fermionic matrices in which the $X's$ appear as derived objects \cite{Banks:2006rx}. However, the main point of this paper is not to give a detailed theory of any particular de Sitter construction, but to explain how the fast-scrambling behavior determines the type of degrees of freedom and how they are connected.

\setcounter{equation}{0}
\section{Comment on Falling}

The following argument is due to Douglas Stanford.  We have seen that the presence of a mass at $x=0$ and $\rho = \R$ is represents a rare fluctuation in which an $m\times m$ block decouples from the rest of the $N\times N$ matrices. Consider how long it takes to return to equilibrium. From the matrix point of view  the answer should  be of order the fast-scrambler-time
\be
\omega^\ast = \log N.
\label{5.1}
\ee

Let us compare the answer \ref{5.1} with the time that it takes for the mass to fall to the stretched horizon of Rindler space. Recall equations \ref{3.1} and \ref{3.2}. The stretched horizon is defined by
\be
\rho = l_s^2
\label{5.2}
\ee
or in terms $x^0, \ x^3,$
\be
(x^3)^2 - (x^0)^2 = l_s^2
\label{5.3}
\ee

The mass $M$ is located at position $x^3 = \R$. It hits the stretched horizon when
\be
(x^0)^2 = \R^2 - l_s^2,
\label{5.4}
\ee
or using \ref{3.1}, when
\be
\sinh^2 \omega = {  \R^2 -l_s^2    \over  l_s^2}.
\label{5.5}
\ee

For $\R >> l_s$ \ref{5.5} becomes
$$
e^{\omega} = {\R   \over l_s } \sim N.
\label{5.6}
$$
or
\be
\omega \sim \log N
\label{5.7}
\ee
which agrees with \ref{5.1}.

\setcounter{equation}{0}
\section{Imprecision}

It's easy to write down matrix models with the symmetry of Rindler space or the static patch of de Sitter space---no doubt much too easy. The models may reflect some of the crude features that we are looking for such as fast scrambling. However, we  have not addressed the much harder question of how we  know if a given model really represents these geometries? For example, is the model consistent with observer complementarity \cite{Parikh:2004wh}. Observer complementarity is the symmetry that relates the observations of two observers who pass out of each other's static patches. From one observers  point of view, another observer gets absorbed into the thermal soup at the first observer's  horizon. From the second observer's point of view nothing special happens at that point, but he sees the first observer fall into the soup.

Mathematically, the issue is whether the
 the additional non-manifest symmetries of the problem can be realized. The symmetry associated with observer complementarity is the non-compact $O(4,1)$ symmetry of de Sitter space, that relates different static patches.  For Rindler space the analogous symmetry it would be the full Poincare group of flat spacetime.

How does one implement such a transformation on the matrix degrees of freedom, which describe only a single static patch? The problem was discussed in \cite{Goheer:2002vf}. The basic idea of that paper is to use Israel's thermofield description of horizons to double the degrees of freedom. The symmetries act on the thermofield double which represents the state on the entire global spatial slice of de Sitter space. However, in \cite{Goheer:2002vf} a no-go theorem was proved: The non-compact $O(4,1)$ symmetry of de Sitter space cannot be realized if the entropy is finite.

The no-go theorem also applies to Rindler space with a regulated area \cite{Goheer:2003tx}.  But in the large $N$ limit in which the area regulator is removed, the entropy diverges and the no-go theorem
disappears. It may be similar to Matrix theory in which the light cone symmetries are manifest but the remaining Lorentz symmetries are not. In Matrix theory the full Lorentz group is only restored in the large $N$ limit. In de Sitter space with a definite $\R$ the full symmetry can only be realized approximately.

These facts are consistent with some recent conjectures about the precision of holographic descriptions in general. In \cite{Harlow:2010my} it was argued that the precision of holographic descriptions is limited by the maximum area that any observer can see in his backward light-cone. Exact holographic duals require that area to be infinite. In de Sitter space the largest area that can be seen by an observer is $\R^2$. In a regulated Rindler space it is $\A$. In neither case do we expect an exact dual.

Both the problem of Observer complementarity and the phenomenological problem having to do with so-called Boltzmann brains  \cite{Dyson:2002pf} are connected  with
exponentially long-time time-scales and Poincare recurrences. New things may happen on such scales which render the problems irrelevant---things like Coleman DeLuccia decays.
On shorter time scales, say of order $H^{-1}$ or $H^{-1}\log{H^{-1}}$ where these problems are not important, de Sitter space makes sense. It is possible that  matrix duals can give a good quantum description of the causal patch for ordinary cosmic time-scales. But, one should not expect such duals to make sense for arbitrary time-scales or to arbitrary precision. Another way to say it is that it may be impossible to distinguish between slightly different duals\footnote{This is a point that has been stressed by Banks and Fischler.}.

\bigskip
\bigskip
\bigskip

To my knowledge, the idea that de Sitter space should have a matrix description first appeared
in \cite{Banks:2006rx}. There is a good deal of overlap with some of the ideas in this paper although the connection with fast scrambling was not mentioned in \cite{Banks:2006rx}. I am grateful to Tom Banks for very interesting discussions.

Recently, at the POTUS conference at Caltech, Erik Verlinde presented some extremely interesting ideas that are also  similar to those in this paper \cite{potus}. My decision to publish this paper was inspired by Erik's talk.

I thank Douglas Stanford for  the observation in Section 5 about the time to fall to the Rindler horizon, as well as other insights.

Finally, over the years I have benefitted from many conversations with Steve Shenker about matrix theories and their connection with string theory and gravity. I am grateful to Steve for sharing his understanding  with me.

\end{document}